\documentclass[aps,pra,twocolumn,showpacs,superscriptaddress]{revtex4-1}

\usepackage{graphicx}				
\usepackage{color}					
\usepackage{bm}					
\usepackage{caption}								
\usepackage{subcaption}				
\usepackage{amssymb}
\usepackage{amsmath}

\begin{document}

\title{Strong-field physics with mid-IR fields}

\author{Benjamin Wolter}
\email{benjamin.wolter@icfo.eu}
\affiliation{ICFO - Institut de Ciencies Fotoniques, Mediterranean Technology Park, 08860 Castelldefels (Barcelona), Spain}

\author{Michael G. Pullen}
\affiliation{ICFO - Institut de Ciencies Fotoniques, Mediterranean Technology Park, 08860 Castelldefels (Barcelona), Spain}

\author{Matthias Baudisch}
\affiliation{ICFO - Institut de Ciencies Fotoniques, Mediterranean Technology Park, 08860 Castelldefels (Barcelona), Spain}

\author{Michele Sclafani}
\affiliation{ICFO - Institut de Ciencies Fotoniques, Mediterranean Technology Park, 08860 Castelldefels (Barcelona), Spain}

\author{Micha{\"e}l Hemmer}
\affiliation{ICFO - Institut de Ciencies Fotoniques, Mediterranean Technology Park, 08860 Castelldefels (Barcelona), Spain}

\author{Arne Senftleben}
\affiliation{Institute of Physics, Center for Interdisciplinary Nanostructure Science and Technology (CINSaT), University of Kassel,
Heinrich-Plett-Strasse 40, 34132 Kassel, Germany}

\author{Claus Dieter Schr\"oter}
\affiliation{Max-Planck-Institut f\"ur Kernphysik, Saupfercheckweg 1, 69117 Heidelberg, Germany}

\author{Joachim Ullrich}
\affiliation{Max-Planck-Institut f\"ur Kernphysik, Saupfercheckweg 1, 69117 Heidelberg, Germany}
\affiliation{Physikalisch-Technische Bundesanstalt, Bundesallee 100, 38116 Braunschweig, Germany}

\author{Robert Moshammer}
\affiliation{Max-Planck-Institut f\"ur Kernphysik, Saupfercheckweg 1, 69117 Heidelberg, Germany}

\author{Jens Biegert}
\affiliation{ICFO - Institut de Ciencies Fotoniques, Mediterranean Technology Park, 08860 Castelldefels (Barcelona), Spain}
\affiliation{ICREA - Instituci\'{o} Catalana de Recerca i Estudis Avan\c{c}ats, 08010 Barcelona, Spain}

\date{\today}

\begin{abstract}

Strong-field physics is currently experiencing a shift towards the use of mid-IR driving wavelengths. This is because they permit conducting experiments unambiguously in the quasi-static regime and enable exploiting the effects related to ponderomotive scaling of electron recollisions. Initial measurements taken in the mid-IR immediately led to a deeper understanding of photo-ionization and allowed a discrimination amongst different theoretical models. Ponderomotive scaling of rescattering has enabled new avenues towards time resolved probing of molecular structure. Essential for this paradigm shift was the convergence of two experimental tools: 1) intense mid-IR sources that can create high energy photons and electrons while operating within the quasi-static regime, and 2) detection systems that can detect the generated high energy particles and image the entire momentum space of the interaction in full coincidence. Here we present a unique combination of these two essential ingredients, namely a 160~kHz mid-IR source and a reaction microscope detection system, to present an experimental methodology that provides an unprecedented three-dimensional view of strong-field interactions. The system is capable of generating and detecting electron energies that span a six order of magnitude dynamic range. We demonstrate the versatility of the system by investigating electron recollisions, the core process that drives strong-field phenomena, at both low (meV) and high (hundreds of eV) energies. The low energy region is used to investigate recently discovered low-energy structures, while the high energy electrons are used to probe atomic structure via laser-induced electron diffraction. Moreover we present, for the first time, the correlated momentum distribution of electrons from non-sequential double-ionization driven by mid-IR pulses.

\end{abstract}

\maketitle
Strong-field physics (SFP) is concerned with the effects related to the interaction of intense electric fields with matter. Here, `intense' means that the electric field strengths are non--negligible compared to the binding fields within matter. Since these field strengths are most easily attained with intense ultrashort laser pulses, most of the effects and observations stem from the interaction with oscillating laser fields. In parallel with the advancement of ultrafast laser science, SFP has developed into a mature field of research that is now capable of tracking electronic and structural dynamics on the atto- to few femtosecond timescales~\cite{Uiberacker2007, Sansone2010, Goulielmakis2010, Smirnova2009, Hockett2011, Zewail2010, Miller2014}. This advance of SFP and attoscience has, in turn, generated an upsurge in the development of ultrafast mid-IR laser sources due to the possibilities when driving strong-field recollision with long wavelengths~\cite{Colosimo2008, Blaga2009, Popmintchev2012, Blaga2012, Dura2013, Ishii2014, Cousin2014}. Mid-IR sources present many benefits compared to the ubiquitous 0.8~$\mu$m radiation of Ti:Sapphire based laser systems. Some of the most relevant to this study are: 1) the unambiguous creation of interaction conditions that are conducive for classical interpretations of experimental results; 2) strong-field recollision at low peak intensity in order to avoid appreciable ground state depletion; and 3) the ability to create high energy recollision electrons for imaging applications. However, long wavelength (ponderomotive) scaling comes at the cost of a dramatically reduced signal~\cite{Colosimo2008,Tate2007,Frolov2008,Austin2012}, which translates into reduced statistics or long data acquisition times. Exacerbating this problem is the fact that traditional electron detection techniques only detect a small fraction of the entire momentum space (see Sec.~\ref{sec:detection_system}). These points are often overlooked but they present a significant roadblock to the advancement of SFP. This is because experiments will eventually become untenable due to the required acquisition times exceeding the stability of the apparatus. Here we present a methodology that overcomes these limitations by combining two tools that are perfectly suited for mid-IR SFP: 1) an intense and high repetition rate optical parametric chirped pulse amplification (OPCPA) based mid-IR (3.1~$\mu$m) source~\cite{Chalus2009} that creates ionization conditions deep in the quasi-static (tunneling) regime and that can generate electron energies above 1~keV; and 2) a reaction microscope (ReMi) detection system that can image the three-dimensional (3D) momentum space of the interaction in full particle coincidence. The apparatus is capable of generating and detecting electron energies that span a six order of magnitude range and thus provides an unprecedented 3D view of strong-field interactions.

The paper is structured as follows: We first discuss relevant aspects of strong-field ionization (SFI) (Sec.~\ref{sec:SFI}) when scaling the driving wavelength of the radiation and not the peak intensity; the metric we use to describe such scaling is the adiabaticity, or `Keldysh', parameter (Sec.~\ref{sec:Keldysh_scaling}). Next, the validity of the dipole approximation is  discussed in the context of wavelength scaling (Sec.~\ref{sec:dipole_approx}). We then highlight the critical role of detection systems with specific attention paid to molecular targets and the importance of measuring all ionization fragments in full coincidence (Sec.~\ref{sec:detection_system}). We proceed to demonstrating the versatility of our methodology by investigating electron recollision (Sec.~\ref{sec:experimental_capabilities}) at the extremes of a 550 eV wide electron spectrum. We highlight two cases: In Sec.~\ref{subsec:low_energy_recollisions} the low-energy part of the spectrum (meV) reveals the recently discovered low energy features for both single and double ions while in Sec.~\ref{subsec:high_energy_recollisions} the high-energy portion (hundreds of eV) is used to extract accurate differential cross-sections (DCSs) over a wide range of recollision energies using laser-induced electron diffraction (LIED). Lastly, we discuss our methodology in the context of the current status and expected direction of SFP research.

\section{Strong-field ionization}
\label{sec:SFI}

\begin{figure}[htb!]
\centering
	\includegraphics[width=0.49\textwidth]{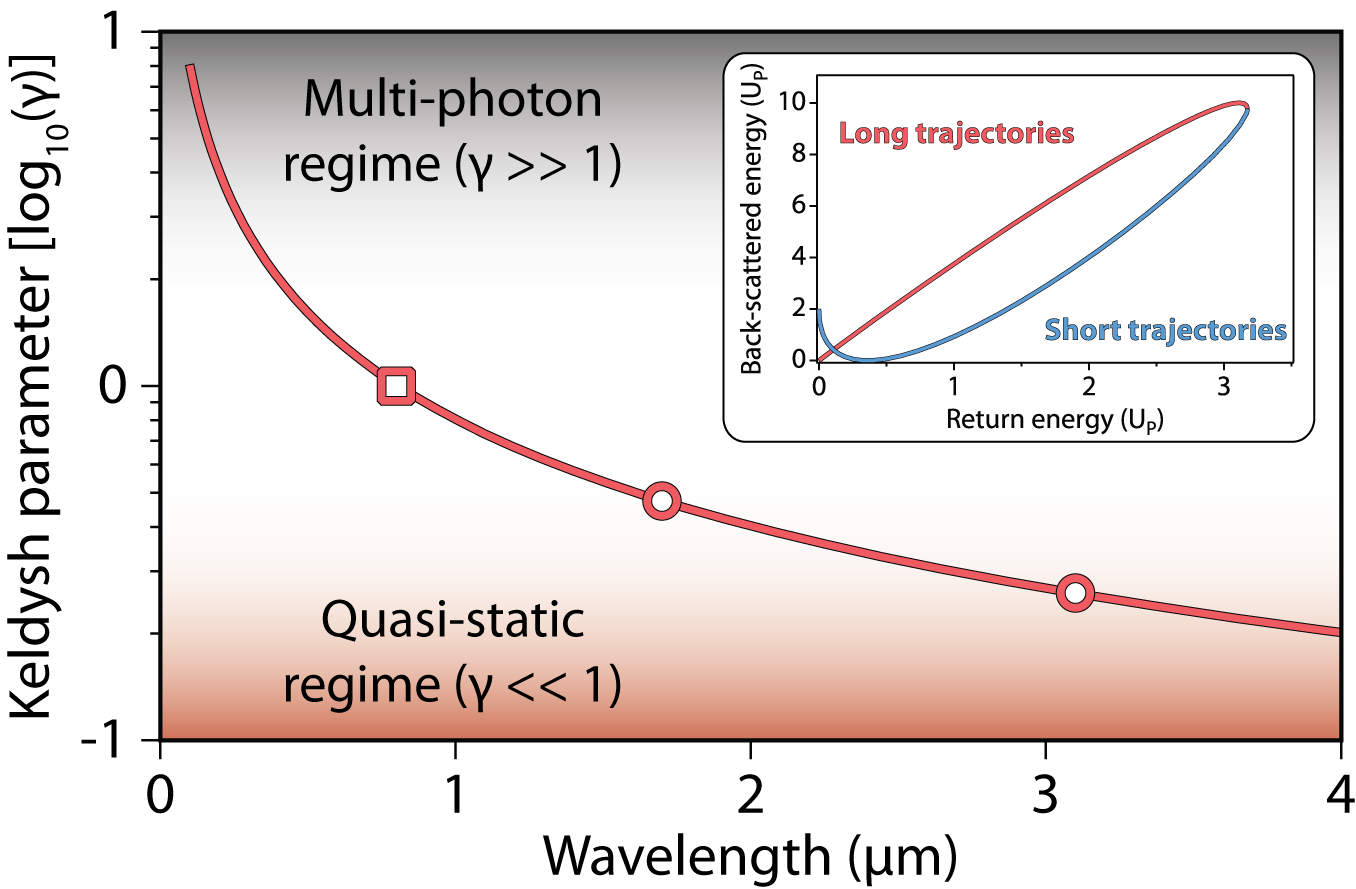}
	\caption{Scaling of the Keldysh parameter versus driving wavelength for xenon atoms ionized with an intensity of $1\times10^{14}$~W\,cm$^{-2}$. The regions of $\gamma\gg1$ and $\gamma\ll1$ are highlighted by grey and red shading, respectively. The ubiquitous Ti:Sapphire wavelength of 0.8~$\mu$m (red square) and our OPCPA outputs (red circles) are presented. The inset shows the kinetic energy $E_{\text{kin}}$ of elastically back-scattered electrons ($\theta_{\text{r}}=180^{\circ}$) versus their return energy $E_{\text{r}}$ (for long and short trajectories) according to the semi-classical `recollision model' in units of the ponderomotive energy $U_{\text{P}}$.}
	\label{fig:keldysh}
\end{figure}

Strong-field interactions can be qualitatively understood by the semi-classical recollision model~\cite{Corkum1993}, which describes the motion of an electron after strong-field induced tunnel ionization. According to this model the electron is emitted close to the peak of the oscillating laser electric field, after which it accelerates within the field before re-colliding with its parent ion roughly three-quarters of an optical cycle later. In the recollision model returning electron trajectories can be distinguished as \emph{long} and \emph{short} depending on the time they spent accelerating in the field before returning to their parent ion. The maximum return energy of $E_{\text{r,max}}$ = 3.17 $U_{\text{P}}$ ($U_{\text{P}}$ being the ponderomotive energy experienced by the detached electron) is achieved when the electron is emitted at a phase of about 17$^{\circ}$ after the field maximum. This results in the electron returning to the ion at about 255$^{\circ}$. Electrons emitted before 17$^{\circ}$ return after 255$^{\circ}$ and are considered long trajectories, while those emitted after 17$^{\circ}$ return before 255$^{\circ}$ and are considered short trajectories.
The model typically 1) neglects depletion of the ground state and Stark shifts; 2) assumes that only one electron participates; and 3) omits the influence of the Coulomb field on the electron's trajectory. Even with such assumptions, numerous potential outcomes of the electron-ion re-collision can be qualitatively described: the returning electron may \textit{re-combine} with the ion and emit a high energy photon in a coherent frequency up-conversion process known as high harmonic generation (HHG); it may \textit{inelastically scatter} of the ion and cause non-sequential double (or multiple) ionization (NSDI); or it may \textit{elastically scatter} off the ion resulting in a final kinetic energy $E_{\text{kin}}$ that is determined by the energy $E_{\text{r}}$ at which it returns to the parent ion and the angle $\theta_{\text{r}}$ at which it scatters. The inset of Fig.\,\ref{fig:keldysh} shows the dependence of $E_{\text{kin}}$ on $E_{\text{r}}$ at $\theta_{\text{r}}=180^{\circ}$ for long and short electron trajectories. A framework to model the ionization step of such strong-field interactions was proposed in Keldysh's seminal work from 1965~\cite{Keldysh1965}. The dimensionless adiabaticity parameter $\gamma$ that is commonly used to describe the interaction conditions was introduced,

\begin{equation}
	\gamma=\sqrt{\frac{I_\text{P}}{2 U_\text{P}}} \sim \frac{1}{\sqrt{I} \lambda}.
	\label{eq:Keldysh}
\end{equation}

\noindent The `Keldysh' parameter $\gamma$ is a function of the ponderomotive energy $U_{\text{P}}\propto I \lambda^2$, where $I$ is the laser peak intensity and $\lambda$ is the central driving wavelength. The scaling of the Keldysh parameter as a function of the driving wavelength is presented in Fig.~\ref{fig:keldysh} for a typical peak intensity of $1\times10^{14}$~W\,cm$^{-2}$ interacting with a xenon (Xe) atom ($I_{\text{P}} \approx 12$~eV). A rapid decrease of the Keldysh parameter can clearly be observed as the driving wavelength is increased from the visible ($\lambda\lesssim0.8~\mu$m) through the near-IR ($0.8~\mu$m~$\leq\lambda\leq3\,\mu$m) and into the mid-IR ($\lambda\geq3\,\mu$m)~\cite{iso}. The circular data points represent the wavelengths that are generated by our OPCPA source. The conventional interpretation of $\gamma$ is as a metric to distinguish between the extremes of two different descriptions of ionization: 1) the so-called `multi-photon' (MP) regime where $\gamma\gg1$ and experimental electron kinetic energy distributions generally feature peaked behavior; and 2) the so-called `tunneling', or to be precise `quasi-static' (QS), regime where $\gamma\ll1$ and the distributions typically exhibit continuous behavior. The case when $\gamma \approx 1$ is known as the `transition regime' and ionization is characterized by features from both MP and QS ionization~\cite{footnote3}. The square data point in Fig.~\ref{fig:keldysh} is within this region and it serves as a reference for the ubiquitous Ti:sapphire radiation at 0.8~$\mu$m. Most SFI experiments performed in the past couple of decades have utilized laser systems based on this technology, therefore theoretical understanding has evolved in this context.

Several motivating factors exist for investigating SFP in the QS regime. The most obvious is the desire to match the assumptions of theoretical models to further our understanding of SFP in general. Another important aspect is the fact that QS conditions are a prerequisite for invoking classical recollision models, which are the cornerstone for developing attosecond science or recollision based time resolved imaging. The requirement for unambiguous conditions was strikingly demonstrated by ionizing atoms with long wavelength radiation that resulted in unexpected structures at low electron kinetic energies~\cite{Blaga2009,Quan2009}. These structures were not predicted by the most widely used approximate theoretical models including the workhorse of the field, the strong-field approximation (SFA). As these models are utilized to interpret many experimental results, including attosecond science experiments, it was worrying that such obvious features could evade prediction and detection for so long. It was not until improvements were made to both the SFA and to classical models that multiple electron re-scattering due to an interplay of the laser and Coulomb fields could be identified as the responsible mechanism~\cite{Faisal2009,Liu2010,Yan2010,Lemell2012,Kastner1,Kastner2,Lemell2013,Guo2013,Lin2014}. Since these initial results a number of other lower energy features have also been observed for both atoms and small molecules during long wavelength investigations within the QS regime~\cite{Wu2012, Dura2013, Pullen2014} and the responsible mechanisms have subsequently been identified~\cite{Wolter2014}. Moreover, the description of the unbound electron being dominated by the laser field within the QS regime is a prerequisite to interrogate inelastic re-scattering processes. The mechanisms behind multiple NSDI connected to electron-electron correlations have been extensively studied and theoretically modeled for near-infrared and high intensity fields, yet questions remain whether the findings hold for different regimes~\cite{Figueira2011}. Finally, the LIED technique, which has recently been utilized to image the structure of both diatomic~\cite{Blaga2012} and polyatomic~\cite{Pullen2014a} molecules, requires invoking the classical recollision model. Here the influence of the strong field is removed and the returning electron wavepacket is treated as a plane wave upon recollision. These motivating factors make it clear that performing SFI experiments in the QS regime is advantageous in many ways.

\section{Scaling of the Keldysh parameter}
\label{sec:Keldysh_scaling}
\begin{figure}[htb!]
	\includegraphics[width=0.49\textwidth]{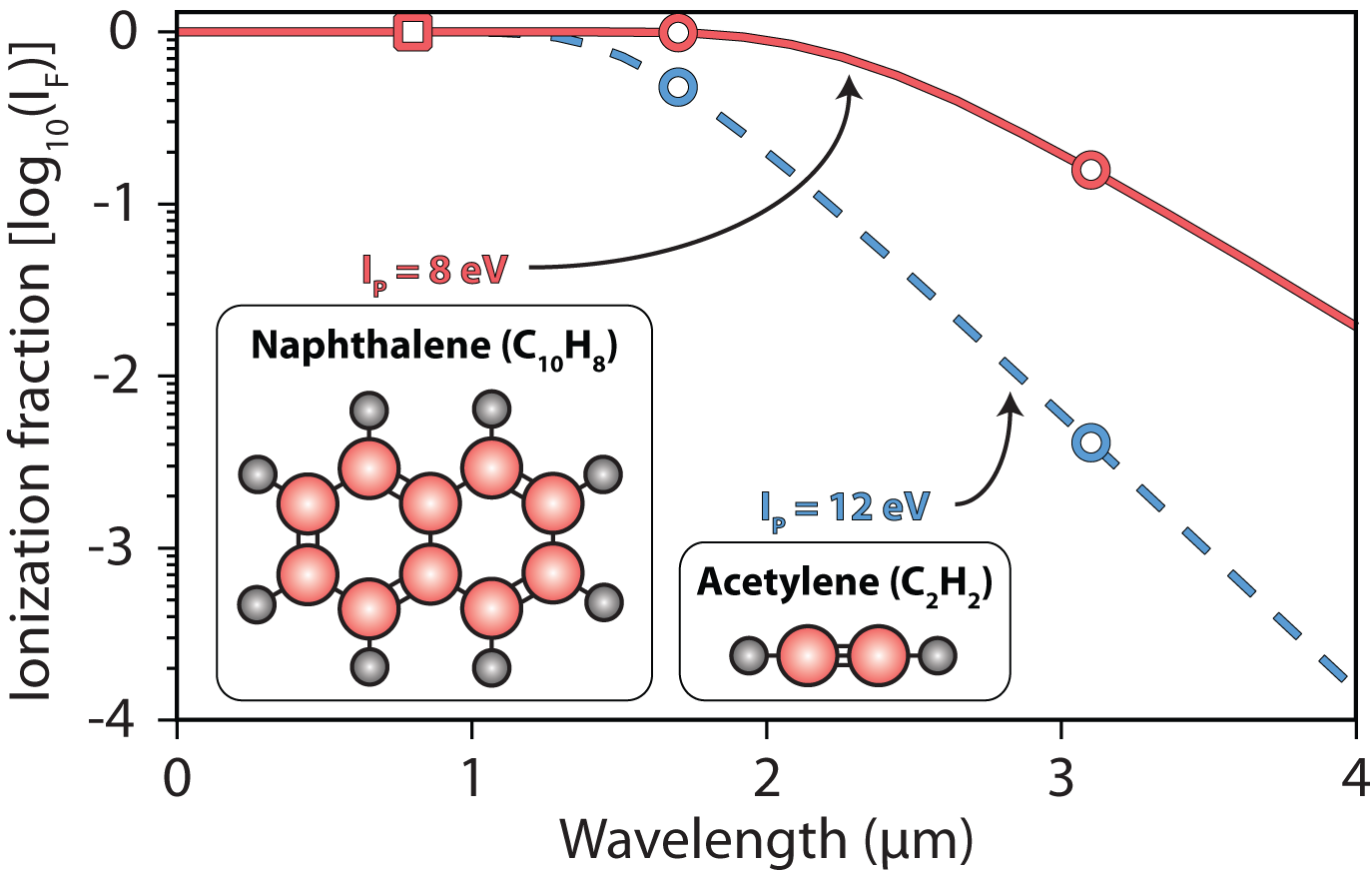}
	\caption{The ionization fractions calculated at the peak of a six cycle pulse for targets with different ionization potentials while keeping the Keldysh parameter constant at $\gamma = 0.3$. The squares represent 0.8~$\mu$m radiation while the circles are the wavelengths of our OPCPA.}
	\label{fig:ionfrac}
\end{figure}

The most straight forward method to reduce $\gamma$ is to increase the laser peak intensity $I$. Unfortunately, from an experimental perspective, this approach is severely constrained by the ionization potential $I_{\text{P}}$ of the target because the ground state population is depleted when the so-called saturation intensity $I_\text{SAT}$ is reached for a given pulse duration. Such intensities cause the fractional amount of re-colliding electrons to reduce due to increased direct ionization on the leading edge of the pulse~\cite{Alnaser2008}. This is detrimental for re-scattering based experiments, therefore it is beneficial for the ionization fraction $I_{\text{F}}$ to be kept much lower than unity. Strictly speaking, as $I_\text{SAT}$ sets the highest effective intensity that a target can experience, it also sets the lower bound on the achievable $\gamma$~\cite{Alnaser2008}. In Fig.~\ref{fig:ionfrac} the ionization fraction at the peak of a six cycle FWHM pulse is illustrated~\cite{Popruzhenko2008} for a target with $I_{\text{P}} = 12$~eV (dashed blue curve) as a function of the laser wavelength while keeping the Keldysh parameter constant at $\gamma=0.3$. Such an ionization potential could represent targets such as Xe or acetylene (C$_\text{2}$H$_\text{2}$), which have ionization potentials of 12.1~ eV and 11.4~eV, respectively. For 0.8~$\mu$m radiation, a peak intensity of $1.1\times10^{15}$~W\,cm$^{-2}$ is required to reach $\gamma=0.3$, which results in complete ionization of the target before the peak of the pulse. Clearly such experimental conditions cannot be used to unambiguously investigate re-scattering phenomena. The same calculation for 3.1~$\mu$m radiation, on the other hand, results in a peak intensity of $7.5\times10^{13}$~W\,cm$^{-2}$ and an ionization fraction of only $4\times10^{-3}$. In fact, it is found that in order to have an ionization fraction $<0.1$ at $\gamma=0.3$, a driving wavelength longer than $2.2~\mu$m is required. This analysis highlights the problem of intensity scaling: it is difficult to generate QS conditions using near-IR radiation without reaching ionization saturation.

Ionization saturation becomes even more problematic when larger polyatomic molecular targets are investigated, as they typically have lower ionization potentials and therefore ionize more easily. To demonstrate the consequences of this aspect, the above calculation was repeated for an ionization potential of about $I_{\text{P}} = 8$~eV (solid red curve in Fig.~\ref{fig:ionfrac}), which could represent a polyatomic molecule such as naphthalene (C$_\text{10}$H$_\text{8}$)~\cite{Dimitrovski2014}. In this case saturation persists until a wavelength of about 2~$\mu$m. At 3.1~$\mu$m, however, the calculated ionization fraction is still only 0.16, which is adequate for the investigation of re-scattering phenomena. Clearly, a more preferable and robust method to achieve QS conditions is to scale the wavelength rather than the intensity~\cite{footnote2}.

\begin{figure}[htb!]
	\includegraphics[width=0.49\textwidth]{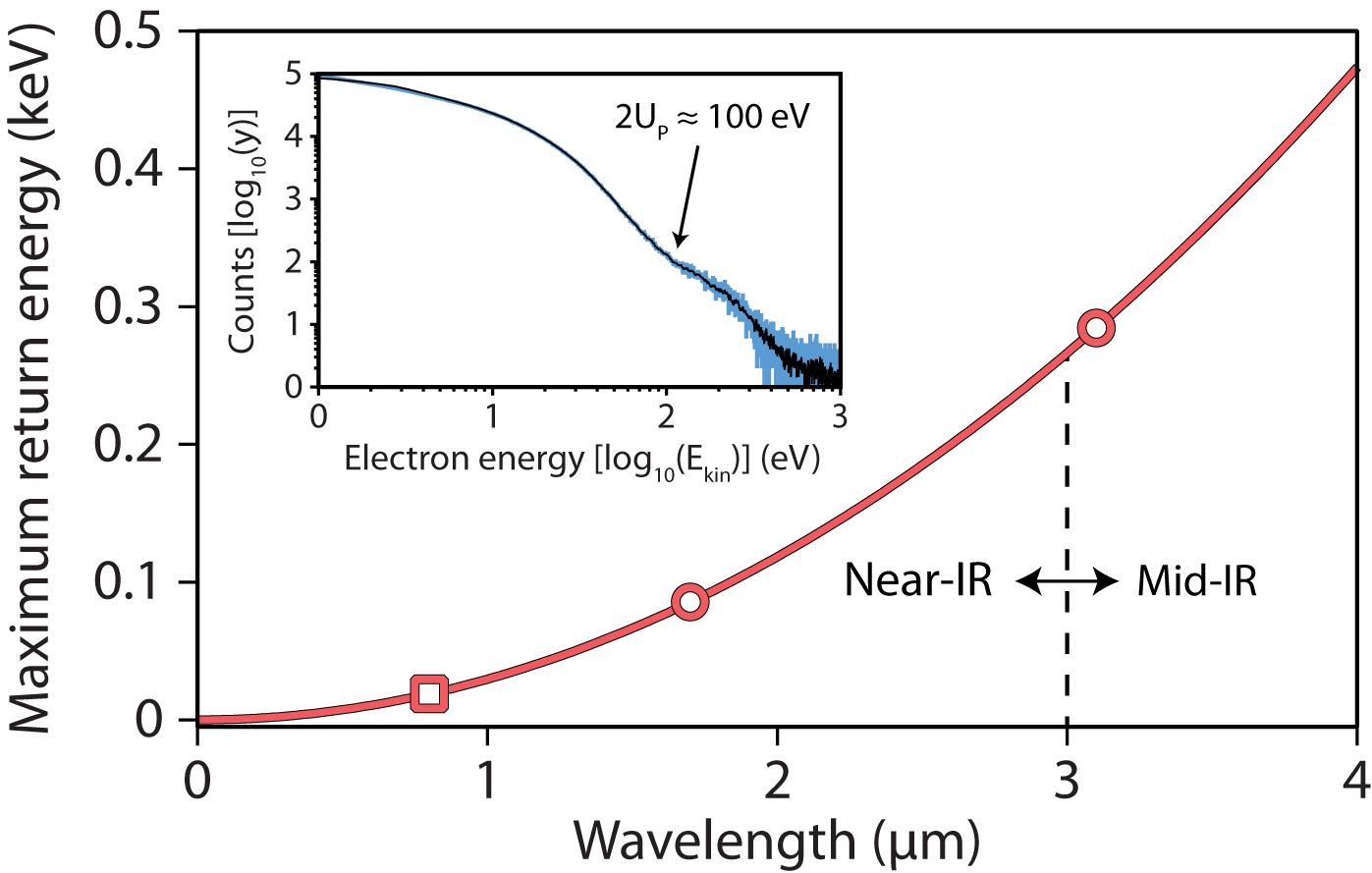}
	\caption{The quadratic scaling of the maximum return energy with wavelength for an intensity of $1.0\times10^{14}$~W\,cm$^{-2}$. The red square represents 0.8~$\mu$m radiation while the red circles are the wavelengths of our OPCPA. The inset shows an electron energy distribution, which extends up to 550~eV, that was measured after ionization of C$_\text{2}$H$_\text{2}$ using the 3.1~$\mu$m output.}
	\label{fig:maxreturn}
\end{figure}

The dependence of $\gamma$ with $\lambda$ exhibits a favourable quadratic scaling of the ponderomotive energy $U_{\text{P}}$. As predicted by the recollision model, the time that the electron accelerates in the field is proportional to the laser field period. For this reason mid-IR radiation, which has a longer optical period, results in much higher energy returning electrons. Increasing the driving wavelength from 0.8~$\mu$m to 3.1~$\mu$m, for example, results in a factor of 15 increase in the ponderomotive energy $U_{\text{P}}$. The recollision model also predicts that the maximum energy that an electron can have upon return to its parent ion is equal to $E_\text{r,max}=3.17\;U_{\text{P}}$. The trend of this maximum return energy versus the driving wavelength is presented in Fig.~\ref{fig:maxreturn} for a typical peak intensity of $1\times10^{14}$~W\,cm$^{-2}$. The square data point shows the maximum energy generated by 0.8~$\mu$m radiation as a reference and the circular data points represent the return energies obtainable with the wavelengths generated by our OPCPA source. Return energies of hundreds of eV can be created by scaling strong-field interaction to mid-IR wavelengths, compared to the tens of eV by standard 0.8~$\mu$m radiation. These much higher energy electrons can be used as a probe of nuclear, rather than electronic, structure and dynamics as is discussed in Sec.~\ref{subsec:high_energy_recollisions}. The inset of Fig.~\ref{fig:maxreturn} shows an experimentally measured electron energy distribution that ranges from $0-550$~eV. In this case the intensity was such that the maximum return energy was $E_\text{r,max} = 175$~eV. Note that the recollision model predicts elastic rescattering to result in overall electron energies of $E_{\text{kin}}\,\lessapprox\,10\,U_{\text{P}}$ (see inset of Fig.\,\ref{fig:keldysh}). In Sec.~\ref{sec:experimental_capabilities} we show that, with the appropriate detection system and analy\-sis techniques, the entire bandwidth of such electron spectra can be utilized to investigate meV features and to extract elastic DCSs at energies of hundreds of eV. High return energies also give access to studying inelastic rescattering processes, such as NSDI, not only from valence electrons but also from inner shells~\cite{Dichiara2012}. One may expect interesting new insights into the multi-electron correlations in many body systems~\cite{Figueira2011} with mid-IR sources. In Sec.~\ref{subsec:low_energy_recollisions_double} we show first results of mid-IR driven double ionization with full coincidence detection.

\section{Validity of the dipole approximation}
\label{sec:dipole_approx}

\begin{figure}[htb!]
	\includegraphics[width=0.49\textwidth]{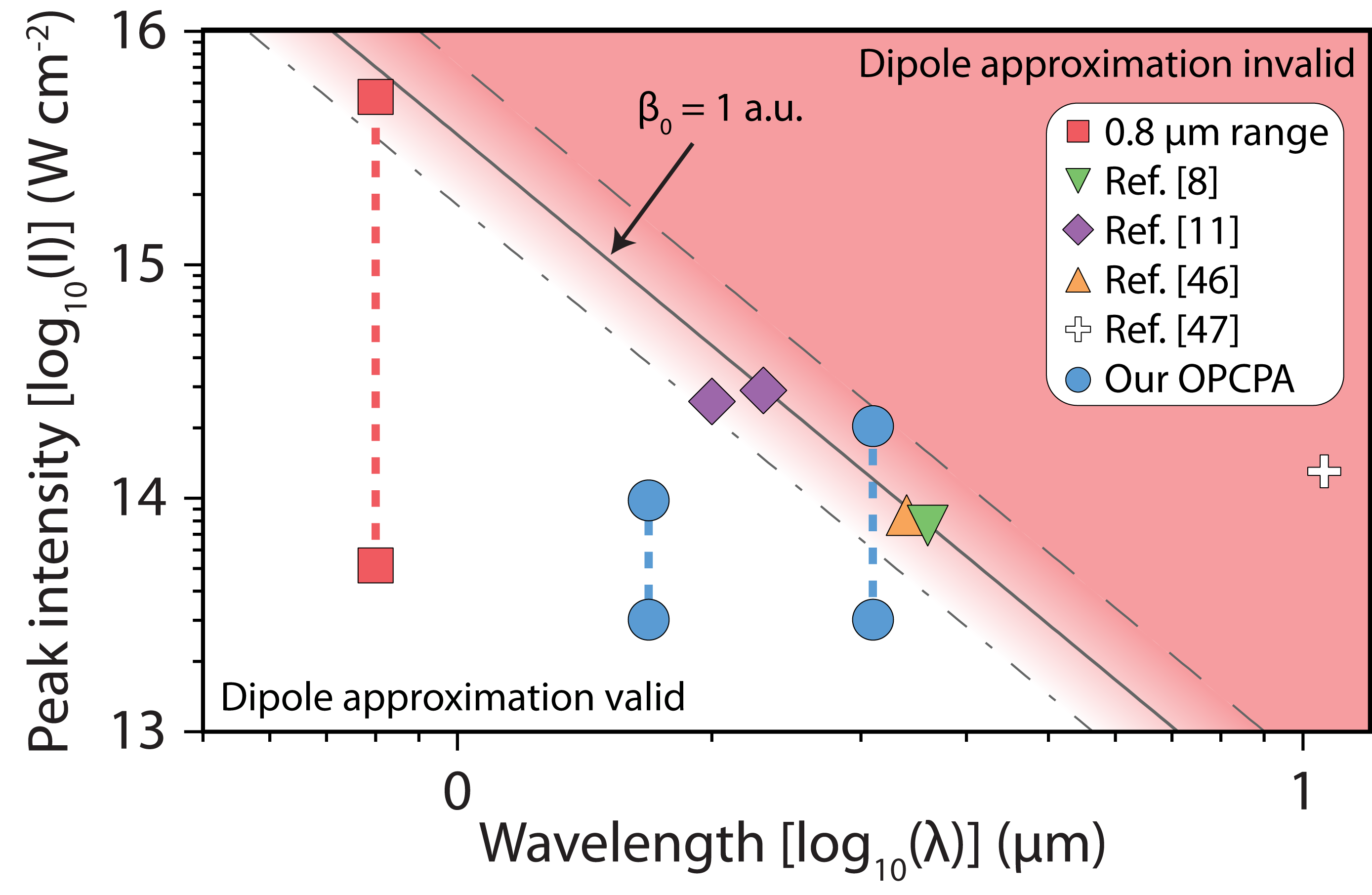}
	\caption{The validity of the dipole approximation as a function of the driving wavelength and peak intensity. The limit where the magnetic field causes the electron motion perpendicular to the laser polarization to become greater than a Bohr radius is given by the solid black line ($\beta_\text{0}=1.0$~a.u.). Also shown are the limits for $\beta_\text{0}=0.5$~a.u. (dot-dashed line) and $\beta_\text{0}=2.0$~a.u. (dashed line). Most 0.8~$\mu$m experiments are performed in the `valid' region (red squares) while many near-IR and mid-IR experiments are close to the limit (Ref.~\cite{Blaga2012} - purple diamonds, Ref.~\cite{Ludwig2014} - orange upwards triangle, Ref.~\cite{Colosimo2008} - green downwards triangle and Ref.~\cite{Xiong1988} - white cross). The range of intensities available to the two outputs of our OPCPA, and therefore the regions over which we can probe the dipole approximation, are presented by blue circles.}
	\label{fig:dipole}
\end{figure}

Many theoretical models used in SFP utilize the dipole approximation~\cite{Reiss1992}, in which the influence of the magnetic field component on the electron motion is assumed to be negligible compared to that of the electric field component. However, when the electron displacement perpendicular to the field polarization direction ($\beta_\text{0}$) becomes comparable to a Bohr radius (i.e. $\beta_\text{0} \approx 1$~a.u.), which can occur at long wavelengths and high intensities~\cite{Reiss2008,Reiss2014}, the validity of the dipole approximation becomes questionable. The limit of $\beta_\text{0} = 1$~a.u. is presented in Fig.~\ref{fig:dipole} (solid black line) for a wide range of experimental driving wavelengths and peak intensities. The regions where the dipole approximation is valid (invalid) are represented by white (red) shading. Also shown are the limits for when $\beta_\text{0} = 2$~a.u. and $\beta_\text{0} = 0.5$~a.u. (dashed and dot-dashed black lines, respectively) to highlight the fact that the validity of the approximation is not defined by a clear threshold.

Most non-relativistic experiments at the standard 0.8~$\mu$m wavelength are performed at intensities between $5\times10^{13}\,-\,5\times10^{15}$~W\,cm$^{-2}$ (red squares and dotted line) where the dipole approximation is generally assumed to be valid. At longer wavelengths the limit of $\beta_\text{0} = 1$~a.u. is reached at lower intensities and a number of experimental measurements with conditions near or above this limit have already been reported in the literature (Ref.~\cite{Colosimo2008} - green down-triangle, Ref~\cite{Blaga2012} - purple diamonds, Ref.~\cite{Ludwig2014} - orange upwards triangle and Ref.~\cite{Xiong1988} - white cross). Also shown are the intensity ranges over which the 1.7~$\mu$m and 3.1~$\mu$m of our OPCPA can be utilized for SFI experiments. The large intensity range of our 3.1~$\mu$m radiation opens up the possibility for investigations across the $\beta_\text{0} < 1$~a.u.~$\rightarrow \beta_\text{0}>1$~a.u. transition. 

\section{On the importance of the detection system}
\label{sec:detection_system}

\begin{figure}[htb]
	\includegraphics[width=0.49\textwidth]{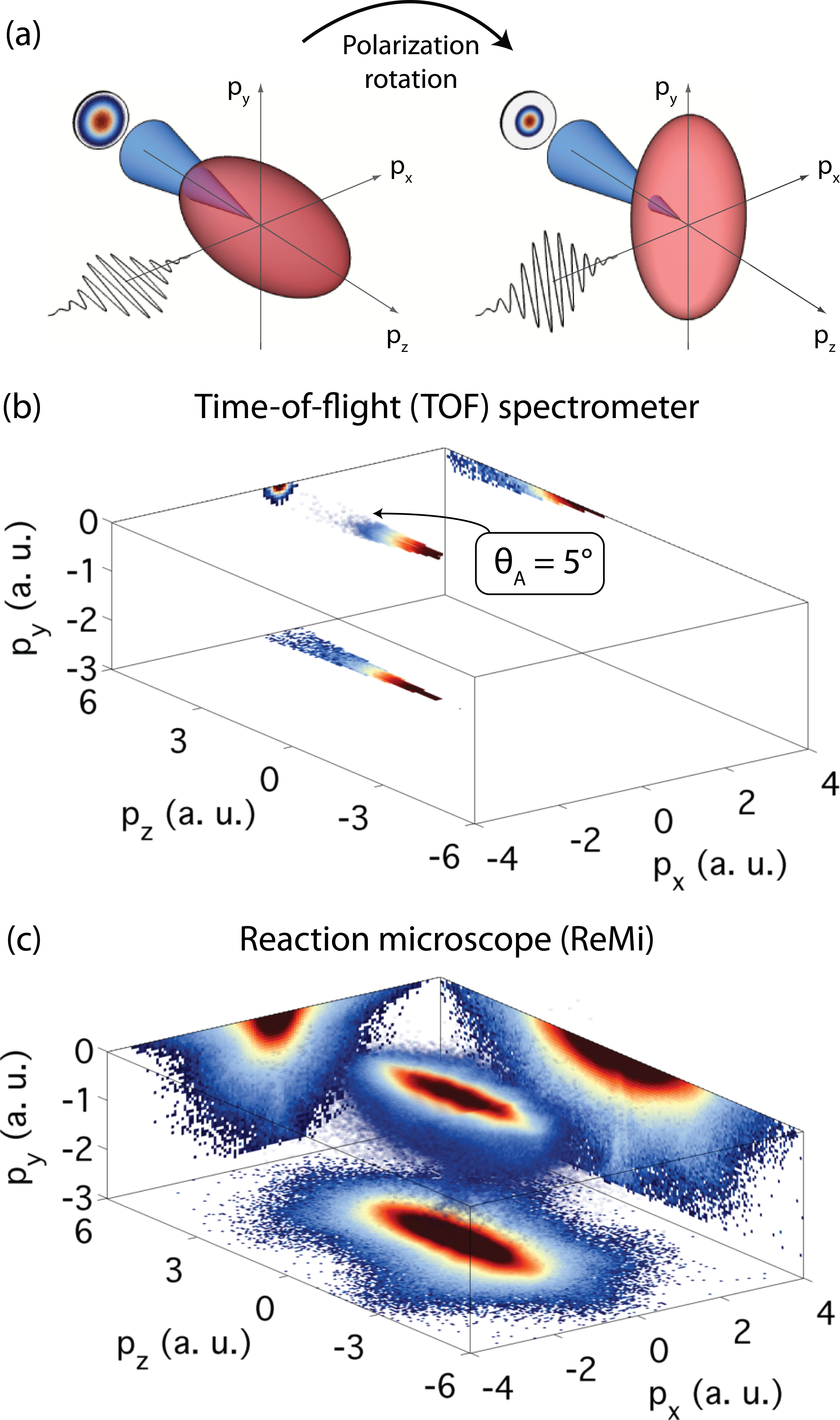}
	\caption{a) Only certain parts of the full 3D momentum distribution (red ellipse) can be measured by a TOF spectrometer, however, the laser polarization can be varied. After ionization of Xe using intense 3.1~$\mu$m radiation, a TOF with an acceptance angle $\theta_\text{A}=5^\circ$ would measure only a small portion (b) of the full 3D distribution collected by the ReMi (c). The highest measured electron momentum for this data was between 6.0\,-\,6.4~a.u., which corresponds to energies between 500\,-\,550~eV.}
	\label{fig:detection_system}
\end{figure}

The interaction between a strong laser field and matter is inherently three dimensional (3D). To help visualize this scenario, a red ellipse is used to approximate a fictitious 3D momentum distribution in Fig.~\ref{fig:detection_system}a. In the case of a linearly polarized field, such as considered here, the momentum distribution is extended more in the direction parallel to the laser polarization (longitudinal direction) than in the transverse directions. An established and still popular method to detect such a distribution is to use a time-of-flight (TOF) spectrometer where the energy of both positively and negatively charged particles can be calculated based on their arrival time after ionization. A typical TOF spectrometer could have an acceptance angle of about $\theta_\text{A}=5^\circ$ (represented by the blue cone), which corresponds to a solid angle that is about 10\% of the full 4$\pi$~sr. For such a detection angle, any particle that has a momentum component perpendicular to the TOF axis that is $\geq9$\,\% of its parallel momentum component (i.e. $p_\perp \geq 0.09\,p_{||}$) will not be detected without rotating the TOF in azimuthal and polar angles about the interaction point. Additionally, any angular information within the acceptance cone is averaged over. Small pinholes can be installed to decrease the angular resolution but this is obviously accompanied by a dramatic drop in count rate. The laser polarization can be continuously adjusted between parallel and perpendicular (as presented in Fig.~\ref{fig:detection_system}a) to the detector axis so that different parts of the momentum cloud can be observed. If this rotation is performed numerous times, a 2D `conical cut' through the momentum distribution can be obtained~\cite{Blaga2012}, however, the integration times required using this method are long. Recent investigations in the QS regime that detected low electron energy ($<1$~eV) structures where $p_\perp \approx p_{||}$~\cite{Wu2012,Dura2013,Pullen2014,Wolter2014} have highlighted the limitations of the TOF measurement technique as without the aid of high resolution 3D detection, these structures evaded observation.

An example of the portion of a 3D momentum distribution that would be measured by a TOF spectrometer is presented in Fig.~\ref{fig:detection_system}b. This data was produced by taking the actual 3D momentum distribution, which was measured by our ReMi after the ionization of Xe gas using intense 3.1~$\mu$m radiation (Fig.~\ref{fig:detection_system}c), and only keeping those counts that were found within the $\theta_\text{A} = 5^\circ$ cone around the laser polarization direction. Only the regions of the distributions with $p_\text{y}<0$ are shown in order to highlight the higher counts near the origin. The projections of the three spatial dimensions are also presented for both measurement methods. The color bar is valid for the 3D cloud but not for the projections as they have each been individually rescaled. Clearly, a TOF spectrometer only collects a small portion of the full electron momentum distribution. Specifically, and as mentioned above, any electrons that have a large transverse component would not be detected. In fact, for an acceptance angle of $\theta_\text{A}=5^\circ$ only about 2\,\% of all electrons are collected. The lack of signal in the TOF projections further highlights these issues. We note that the ReMi and TOF spectrometer distributions should be compared in a qualitative manner only as it is possible for TOF spectrometers to have a different detection angle, to be utilized in a stereo configuration, and to be used for varying laser polarization directions.

\begin{figure}[htb!]
	\includegraphics[width=0.49\textwidth]{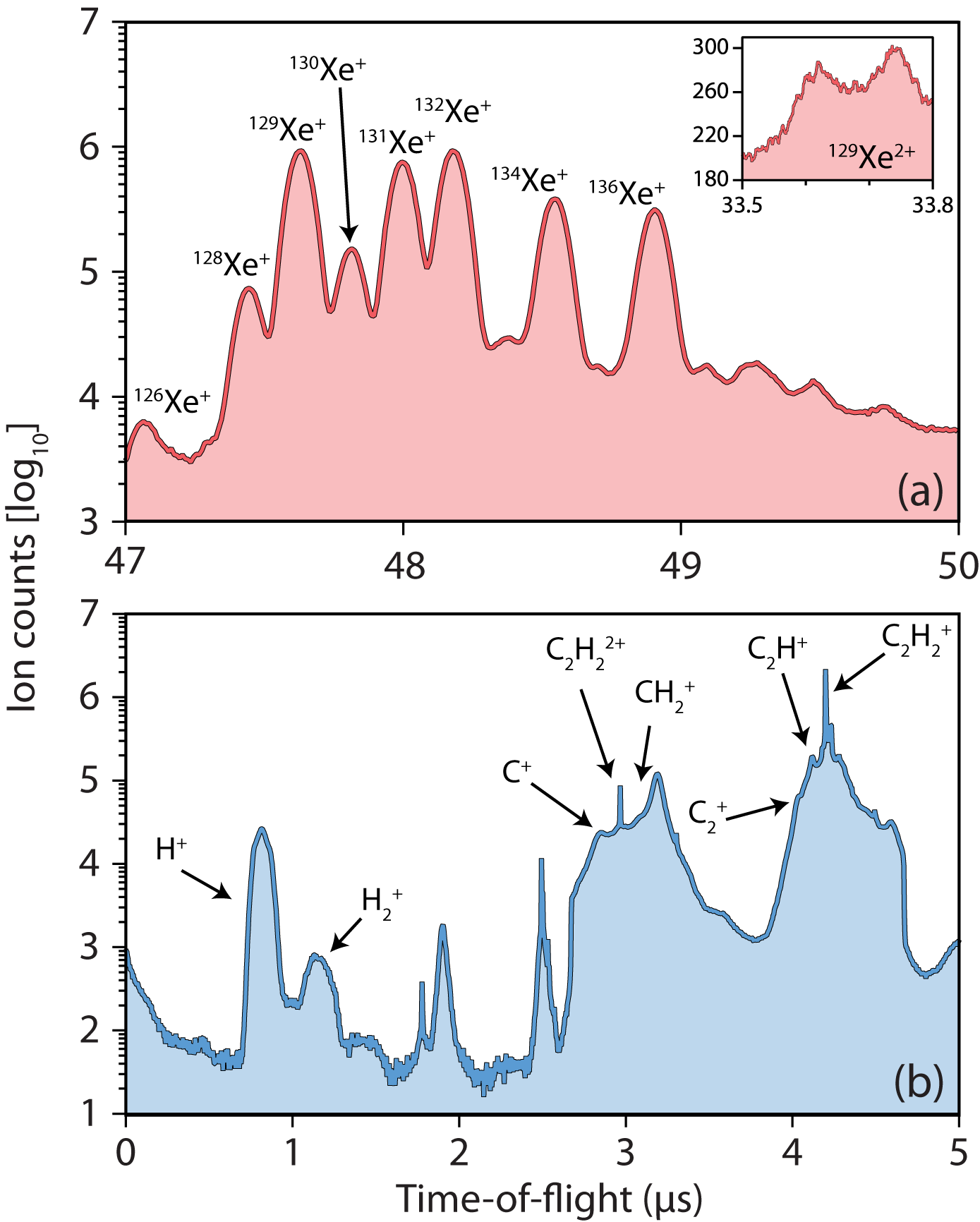}
	\caption{a) The measured ion TOF spectrum after the ionization of Xe using the 3.1~$\mu$m output. The individual Xe isotopes can be accurately resolved when low $|\vec{E}|$ and $|\vec{B}|$ extraction fields are used. The inset shows TOF information corresponding to $^{129}$Xe$^{2+}$. b) The same as in (a) except for C$_\text{2}$H$_\text{2}$ and with high extraction fields. Many molecular fragments are observed with each one having its own corresponding electrons.}
	\label{fig:Xe_C2H2_TOFs}
\end{figure}

In addition to momentum distributions being intrinsically three dimensional, electrons may be detected that originate from all atomic and molecular fragments in the interaction region. This can include ions and isotopes of the target of interest, fragments generated from dynamical processes such as dissociation and Coulomb explosion, as well as background gases. The only way to separate out target specific electronic information is to correlate the detected electrons with their ionic partners \cite{Moshammer1994,Frasinski1989}. An example of an ionic TOF detected after the ionization of Xe by intense 3.1~$\mu$m radiation is presented in Fig.~\ref{fig:Xe_C2H2_TOFs}. The singly ionized stable and long lived isotopes of Xe are resolved and their relative abundances agree with the known values~\cite{CRC}. Using a two particle coincidence condition, the electrons corresponding to the Xe$^+$ ions can be isolated from the rest of the electron background. In Sec.~\ref{subsec:low_energy_recollisions} we perform such an analysis to investigate the low momentum part of the electron spectrum and we present why many of the features evaded detection with TOF spectrometers.

Statistics permitting, higher order particle coincidences are possible for multiple ionized ions. For example, the two peaks~\cite{footnote1} corresponding to the doubly ionized $^{129}$Xe isotope are visible in the inset of Fig.~\ref{fig:Xe_C2H2_TOFs}. As this is a double ion, it has two corresponding electrons. Using a three particle coincidence condition, we can isolate the two electrons corresponding to each double ion. In Sec.~\ref{subsec:low_energy_recollisions} we present the results of such analysis and show, for the first time, an electron-electron correlation map resulting from NSDI triggered by intense mid-IR radiation.

The importance of coincidence detection becomes even more vital when molecular targets are investigated as they can undergo fragmentation upon SFI. In Fig.~\ref{fig:Xe_C2H2_TOFs}b the ionic TOF observed after the ionization of C$_\text{2}$H$_\text{2}$ using similar laser parameters is presented. Clearly many fragmentation channels can be observed. Without coincidence detection, electrons not associated with the fragment of interest become an unwanted background signal that can detrimentally affect data analysis or cloud interpretation. This point is discussed in detail with respect to performing LIED on polyatomic molecular targets in Ref.~\cite{Pullen2014a}. As SFP continues to progress from atomic and simple homonuclear diatomics towards more complex molecular targets, coincidence detection will become increasingly vital to extract specific information within a myriad of competing processes.


There are currently two experimental methods that can achieve 3D detection in full coincidence: Firstly, velocity map imaging (VMI)~\cite{Eppink1997,Takahashi2000, Laksman2013} is a technique where particles with the same mass and initial velocity are mapped onto the same position on a two dimensional (2D) detector. A VMI can provide electronic and ionic kinetic energies and angular distributions with full 4$\pi$~sr collection whilst achieving energy resolutions on the percent level~\cite{Ghafur2009}. Most VMI apparatuses require cylindrical symmetry to transform the detected projection into a slice through the centre of the 3D momentum distribution, however, in special implementations this is not necessarily the case~\cite{Gebhardt2001, Whitaker2000,Vrakking2001, Dribinski2002}. Moreover, VMIs typically only operate in full particle coincidence in special circumstances~\cite{Rolles2007} rather than by default. The second option is to use a ReMi~\cite{Moshammer1996,Dorner2000,Ullrich2003}, as is this case in this manuscript, which is based on cold-target recoil ion momentum spectroscopy (COLTRIMS). The arrival time and position of all charged particles is detected in order to reconstruct the full 3D doubly differential cross-section of the interaction. The operating principle of a ReMi is described in Sec.~\ref{sec:experimental_apparatus} A.

\section{Experimental system}
\label{sec:experimental_apparatus}

\begin{figure*}[htb]
	\includegraphics[width=0.98\textwidth]{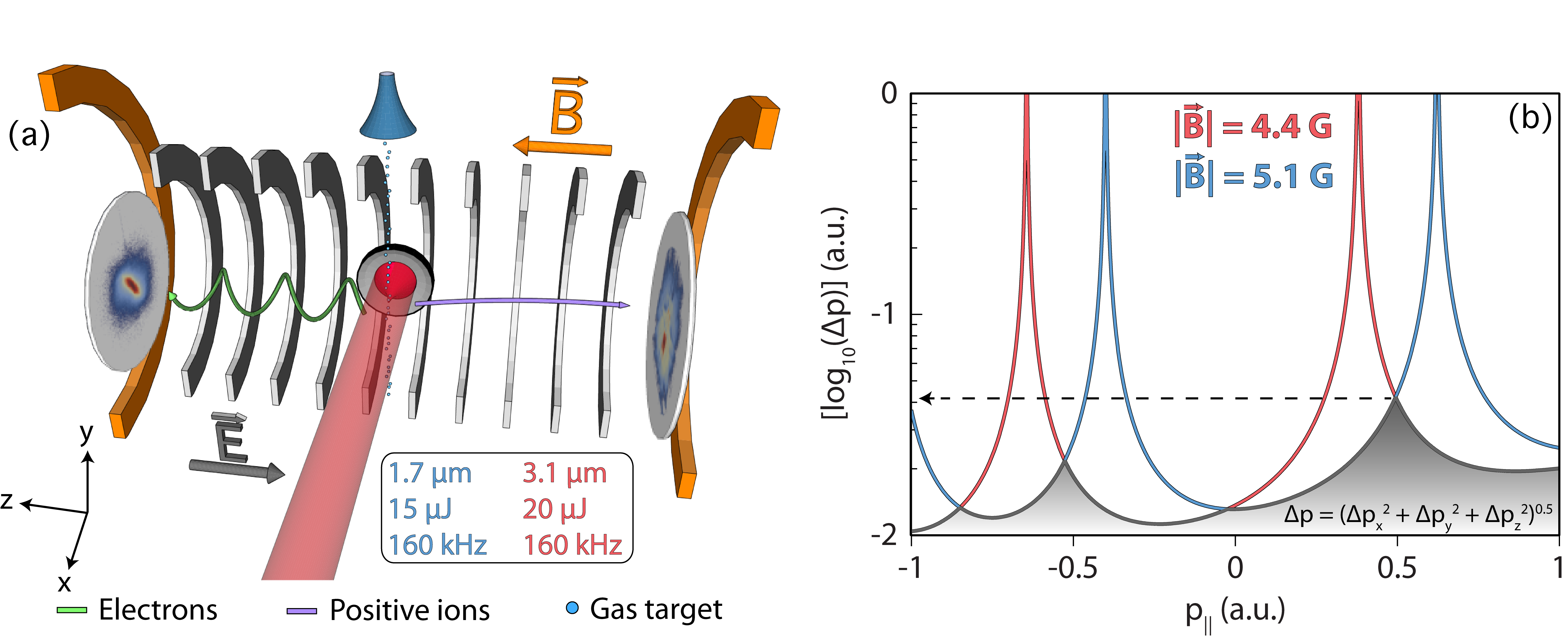}
	\caption{a) The experimental apparatus used to study SFP at long wavelengths. The operating principle of the ReMi is described in the text. The 160~kHz OPCPA has two outputs that can be focussed into a supersonically expanding gas jet to induce SFI. The full 3D momentum distribution of the interaction can be mesaured. b) The momentum resolution of the ReMi as a function of $p_{||}$ for two low values of the magnetic field. The `nodes' can be eliminated by stitching the results of multiple measurements with different values of $\vec{B}$ together.}
	\label{fig:experimental_apparatus}
\end{figure*}

Our experimental system consists of a high repetition rate mid-IR OPCPA source with long term power and carrier-envelope phase (CEP) stability in combination with a ReMi detection system. The mid-IR source provides conditions in the QS regime while the ReMi images the entire 3D momentum distribution of the interaction in full coincidence. Overall, the system is currently capable of creating and detecting electron energies over a six order of magnitude range and it provides the perfect conditions to investigate and utilize mid-IR SFI in the QS regime.

\subsection{Reaction microscope (ReMi)}
\label{subsec:ReMi}

The functional principle of the ReMi detection system is as follows. The target gas is supersonically expanded into the ultra-high vacuum ($10^{-11}$~mbar without gas load) and is subsequently skimmed in two successive stages before reaching the ReMi. The expansion of the molecular beam thus features a decreased thermal motion of the particles and internal temperatures to the milikelvin range are feasible. Upon SFI, a combination of static electric ($\vec{E}$) and magnetic ($\vec{B}$) extraction fields direct all charged fragments to two opposing large-scale and position sensitive micro channel plate (MCP) detectors. The homogeneous electric field is created by a stack of metal electrodes inside the ReMi while the magnetic field is created by high current external coils. In Fig.~\ref{fig:experimental_apparatus}a the negative and positive particles created during the ionization event are represented by green and purple arrows, respectively. The 3D momentum vectors (with the $x$-axis along the laser propagation direction, the $y$-axis being anti-parallel to the direction of the gas jet and the $z$-axis along the direction of the static fields) and hence the kinetic energy of the particles can be calculated from the measured TOF and the position of impact. The choice of the electric and magnetic fields define both the momentum range that can be examined in the experiment and the achievable momentum resolution. By choosing relatively low fields, high momentum resolutions can be achieved, as presented in Fig.~\ref{fig:experimental_apparatus}b. Here the total momentum resolution ($\Delta p = \sqrt{\Delta p_\text{x}^2 + \Delta p_\text{y}^2 + \Delta p_\text{z}^2}$) has been calculated~\cite{Ullrich2003} as a function of the longitudinal momentum ($p_{||} = p_\text{z}$) while $|\vec{E}| = 1.3$~V/cm is kept constant for both $|\vec{B}| = 4.4$~G (red curve) and $|\vec{B}| = 5.1$~G (blue curve). The observed discontinuities for both magnetic fields are due to electrons with integer multiples of the cyclotron frequency hitting the same position on the detector. While these so-called `nodes' seem inconvenient at first, they turn out to be useful as they serve as an accurate and straight forward measurement of $\vec{B}$. Moreover, is it easy to remove these regions by taking two measurements with different magnetic fields and combining the areas with the highest momentum resolution. The results of this technique are represented by the grey shaded region in Fig.~\ref{fig:experimental_apparatus}b where $\Delta p \leq 5\times10^{-2}$~a.u. for all values of $p_{||}$. In terms of energy these values correspond to detection resolutions of several meV. It is straightforward to further reduce the electric and magnetic fields to achieve even better momentum resolution. By choosing higher values of $|\vec{E}|$ and $|\vec{B}|$ it is possible to detect re-scattered electrons with kinetic energies of up to 1~keV. The magnitude of $\vec{E}$ controls the maximum detectable longitudinal electron momentum component ($p_{||}$) while the maximum transverse component ($p_\perp = \sqrt{p_\text{x}^2 + p_\text{y}^2}$) is dependent on the magnitude of $\vec{B}$. Figure~\ref{fig:detection_system}c presents a logarithmically scaled 3D momentum distribution detected under these high static field conditions ($|\vec{E}|=51$~V/cm and $|\vec{B}|=39$~G) where electron energies of $\geq500$~eV were detected.

\subsection{Mid-IR OPCPA system}
\label{subsec:OPCPA}

The OPCPA system~\cite{Thai2011,Hemmer2013a} is based on a multi-color fiber frontend with difference frequency generation (DFG) between the two outputs supplying broadband mid-IR pulses at 3.1~$\mu$m center wavelength. The DFG stage is a crucial component of the OPCPA as it provides passive carrier-envelope phase (CEP) stability optically. The intrinsic CEP stabilization is a decisive advantage over electronically stabilized systems which manifests itself in unsurpassed long time stability of our system. The system provides both 1.7~$\mu$m and 3.1~$\mu$m radiation. At 3.1~$\mu$m, pulse durations below three optical cycles FWHM are achieved (32~fs = 2.9 optical cycles)~\cite{Hemmer2013b}. The 3.1~$\mu$m output is passively CEP stable and a stability of 250~mrad rms over 11 minutes has been measured~\cite{Thai2011}. Without any additional effort, pulse durations of 100 fs are reached for the signal at 1.7 $\mu$m. The 15~$\mu$J and 20~$\mu$J pulse energies of the 1.7~$\mu$m and 3.1~$\mu$m wavelengths, respectively, are focused with an on-axis, gold-coated paraboloid (focal length of 50~mm) into the gas jet of the ReMi. Intensities in the range of $10^{14}$~W~cm$^{-2}$ are readily achieved for both wavelengths. 

\subsection{The importance of repetition rate and stability}

A major challenge in performing SFP experiments with long wavelength sources is quantum diffusion \cite{Watson1997}. Here the electron wavepacket experiences larger longitudinal and transverse spreading due to the longer excursion time at increased wavelength. Consequently, the larger extent of the recolliding wavepacket results in a smaller recombination or recollision cross-section which manifests in reduced count rate when driving SFP at longer wavelength~\cite{Tate2007,Frolov2008}. Experimental evidence shows that the ratio of re-scattered to directly emitted electrons scales roughly as $\lambda^{-4}$~\cite{Colosimo2008}. 

\begin{figure}[htb!]
	\includegraphics[width=0.49\textwidth]{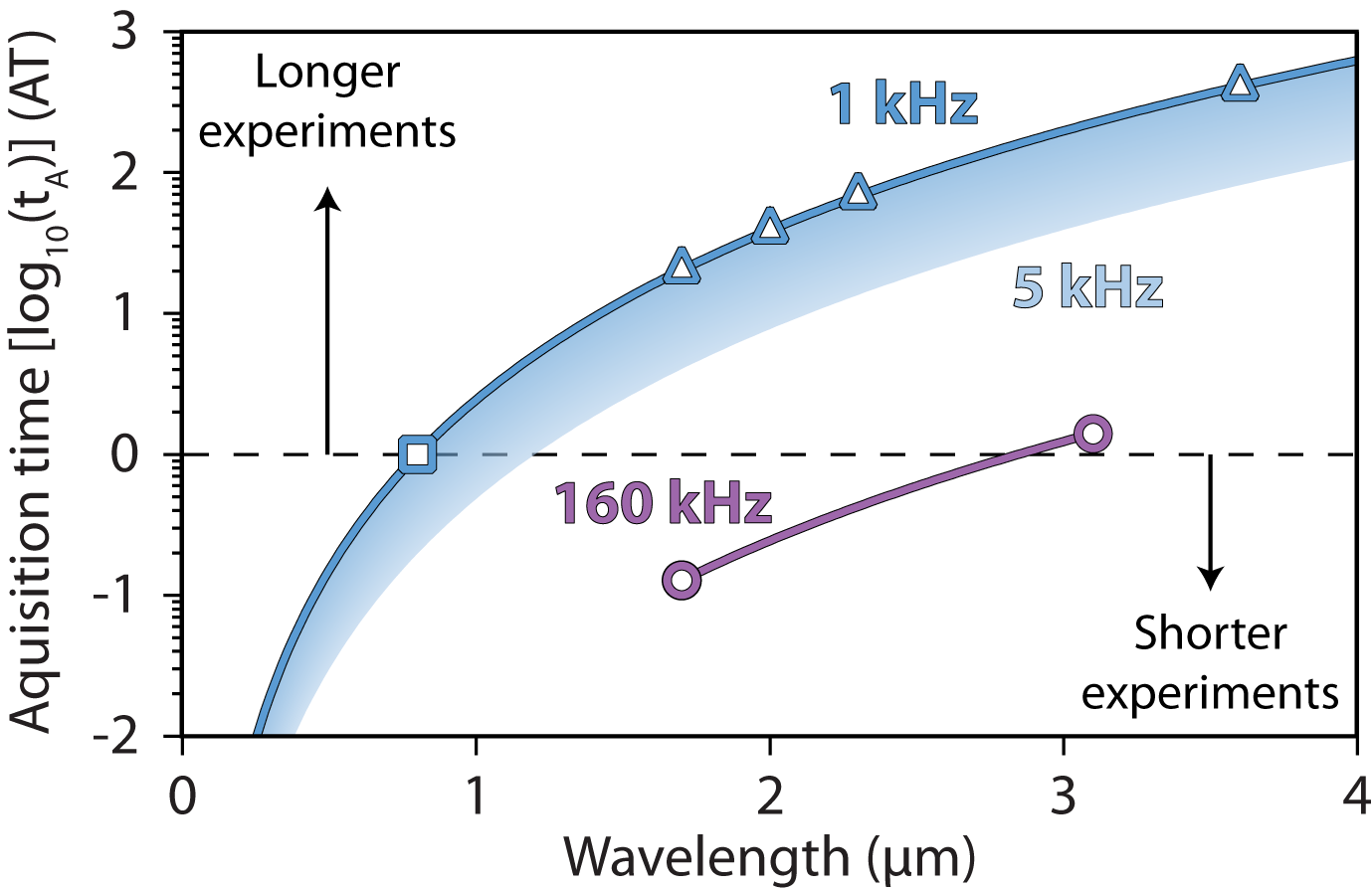}
	\caption{The number of `acquisition times' (as defined in the text) required for a range of driving wavelengths after accounting for the $\lambda^{-4}$ scaling of electron re-scattering. Typical 1~kHz systems based on Ti:Sapphire technology (Ref.~\cite{Blaga2012} - blue triangles) are limited as total acquisition times become untenable. Even the expected repetition rate upper limit of 5~kHz isn't fast enough to compensate for the decrease of signal (shaded region). Using a 160~kHz system such as our OPCPA, on the other hand, allows much shorter acquisition times (purple circles).}
	\label{fig:dataacq}
\end{figure}

We choose to highlight this crucial point for experimental investigations in Fig.~\ref{fig:dataacq} by defining an arbitrary standard unit of `acquisition time' (AT) to be the time required to reach a certain signal-to-noise ratio of re-scattered electrons when using a standard 1~kHz, 0.8~$\mu$m radiation source (dashed black line). Using the above mentioned $\lambda^{-4}$ scaling and assuming all other experimental parameters stay constant, we can estimate the number of standard ATs required to reach the identical signal-to-noise ratio as a function of driving wavelength. The results of this calculation are presented in Fig.~\ref{fig:dataacq} for 1~kHz sources (blue curve). The triangular data points represent the wavelength of some sources that have recently been used to study electron re-scattering~\cite{Colosimo2008, Blaga2009, Blaga2012}. The required ATs for identical signal-to-noise scale rapidly with wavelength; compared with 0.8~$\mu$m radiation we predict a one and two orders of magnitude increase for 1.4~$\mu$m and 2.5~$\mu$m radiation, respectively. At a wavelength of 3.6~$\mu$m the required AT is over 400. We would like to put the arbitrarily defined ATs and the quoted numbers into perspective: if a standard AT is equal to one hour then at 3.6~$\mu$m the new experimental time would be over 17 days. Clearly, the unfavorable wavelength scaling of SFP must be addressed to ensure continual advancement of the field.

An obvious solution to this dilemma is to increase the repetition rate of the driving laser~\cite{Chalus2008,Chalus2009,Furch2013,Matyschok2013}. Ti:Sapphire based laser systems~\cite{Colosimo2008,Blaga2009,Quan2009,Wu2012} are typically limited to a few kHz; the range of acquisition times that can realistically be achieved with such systems is represented by the blue shaded region in Fig.~\ref{fig:dataacq}. Mid-IR sources at much higher repetition rates, i.e. above several tens of kilohertz, can be realized with a range of techniques but so far only OPCPA has provided a viable solution. High average power fiber laser sources are operational, but additional down conversion into the mid-IR carries a dramatic penalty in pulse energy. Moreover, few-cycle durations and CEP stability are not possible with such sources at high intensities ($\geq10^{13}$~W\,cm$^{-2}$). Another approach could be based on passive enhancement cavities~\cite{Pupeza2010} in the mid-IR but such apparatuses are extremely challenging and are yet to be implemented outside of the near-IR. Due to these technical problems we have devised a radically new platform~\cite{Chalus2008,Chalus2009} (summarized in Sec.~\ref{subsec:OPCPA}) that leverages the concept of OPCPA to achieve {\it the combination} of mid-IR wavelength (3.1~$\mu$m), few-cycle pulse duration ($\geq$ 2.9 cycles) with CEP stability, SFP relevant intensities ($\leq 3\times10^{14}$ W cm$^{-1}$), 160 kHz repetition rate and long term stability ($\leq 1\%$ rms over 4.5 hours)~\cite{Hemmer2013a}. Additionally, parasitically generated wavelengths such as the OPCPA's signal at 1.7~$\mu$m can be used for further frequency down-converison. When the above `acquisition time' analysis is applied to the two wavelengths of our system it is found that the required experimental time is actually an order of magnitude less than a standard AT for 1.7~$\mu$m  and only 40\% longer for 3.1~$\mu$m (the purple circles in Fig.~\ref{fig:dataacq}). Combining these points with the fact that the mid-IR source can actually reach a $\gamma<0.2$, it is clear that mid-IR OPCPA solves the problem of studying SFP at longer wavelengths.

\section{Experimental capabilities}
\label{sec:experimental_capabilities}

We now demonstrate the capabilities of our approach by investigating strong-field electron recollision at low and high energies. We choose Xe as an example for which we measure a $550$~eV wide kinetic energy spectrum after  ionization with intense 3.1~$\mu$m radiation. By optimising the electric and magnetic guiding fields of the ReMi, as discussed in Sec.~\ref{sec:experimental_apparatus}, we  investigate different parts of the spectrum. With low fields we resolve meV features at energies below 1~eV while with high fields we collect all ejected electrons up to energies of hundreds of eV.

\subsection{Low energy recollisions}
\label{subsec:low_energy_recollisions}

As already described briefly in Sec.~\ref{sec:SFI}, experimental investigation of SFI with longer wavelength laser sources ($\lambda \geq 1.5~\mu$m) revealed unexpected behavior in the observed electron energy spectra that deviated from SFA predictions~\cite{Blaga2009,Quan2009}. In addition to the well-known form of the kinetic energy spectrum and the drop in yield beyond directly emitted electrons, a peaked structure at low energies ($1\,-\,3$~eV) was observed. This feature, dubbed the low-energy structure (LES), was found to result from the interplay between the Coulomb binding potential and the laser electric field~\cite{Faisal2009,Liu2010,Yan2010}. Additional features have been observed at even lower energies and were dubbed very low energy structures (VLES)~\cite{Wu2012}. In these cases the description of the interaction using SFA based models seemed to be missing crucial aspects of the post ionization behavior of the electron in the laser field. It was found that multiple `revisits' of the electron close to the ionic core, and therefore within the influence of the Coulomb field, had the affect of bunching the oscillating electrons in momentum space~\cite{Lemell2012,Kastner1,Kastner2,Lemell2013,Guo2013,Lin2014}. It is interesting to note that classical trajectory calculations that ignore the Coulomb potential were able to reproduce the LES features~\cite{Becker2014}, however, the Coulomb potential markedly enhances the effect. From an experimental point of view it is important to remark that the LES was originally measured using a TOF spectrometer, which, as explained in Sec.~\ref{sec:experimental_apparatus}, only detects a small fraction of the entire momentum distribution. This restriction in detection translated into incomplete coverage of the momentum space, thereby missing a large fraction of information from the observed effects. A comprehensive picture of the observed effects was gained from our full 3D and high resolution momentum measurements with a ReMi~\cite{Dura2013,Pullen2014}. In these measurements, an additional feature near zero momentum was observed and called the zero energy structure (ZES). Recently, we performed a combined experimental and theoretical investigation \cite{Wolter2014} in which the origins of all three features were quantitatively confirmed to be due to two-dimensional Coulomb focusing (for the LES and VLES) and post-ionization recapture of electrons into high-lying Rydberg states with subsequent field ionization (for the ZES). The latter mechanism is also known as frustrated tunnel ionization~\cite{Nubbemeyer2008,Landsman2013}.

\subsubsection{Singly ionized Xe}
\label{subsec:low_energy_recollisions_single}
\begin{figure}[htb!]
	\includegraphics[width=0.49\textwidth]{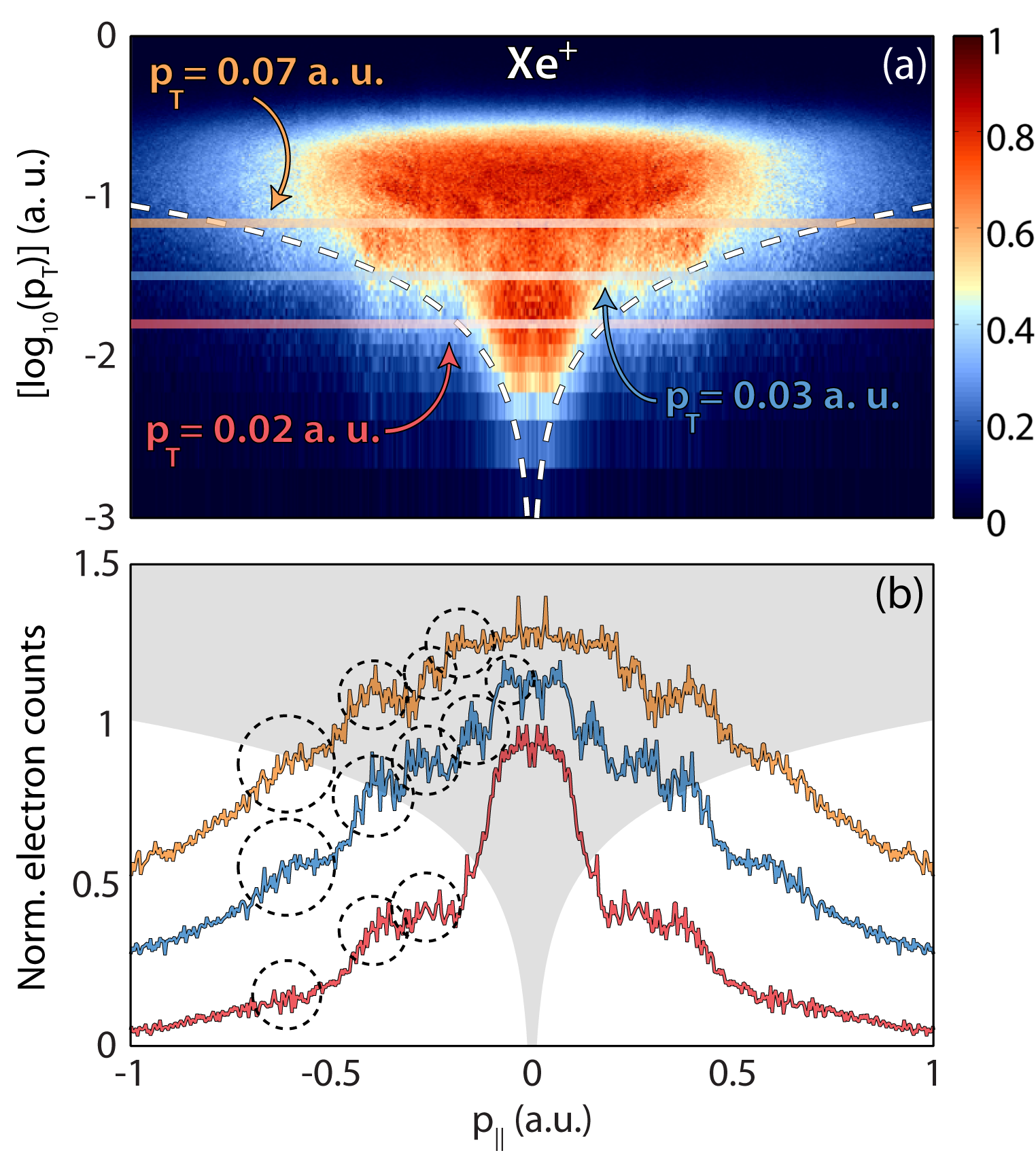}
	\caption{a) The measured electron momentum distribution corresponding to Xe$^+$ after interacting with an intense 71~fs, 3.1~$\mu$m pulse with a peak intensity of $I=4\times10^{13}$~W\,cm$^{-2}$ ($\gamma=0.4$). The data has been symmetrized with respect to $p_\parallel=0$. Line-outs at three values of $p_\perp$ are presented in (b) to highlight the different LES orders and the VLES (black dashed circles). The white dashed line in (a) represents the acceptance cone of a TOF spectrometer with $\Theta_\text{A}=5^\circ$, which roughly precludes the detection of the features within the grey shaded region in (b).}
	\label{fig:Xe1p}
\end{figure}

Figure~\ref{fig:Xe1p} presents kinematically complete measurements of single ionization of Xe using 3.1~$\mu$m radiation at a peak intensity of $I=4\times10^{13}$~W\,cm$^{-2}$ and a pulse duration of 71~fs (7 optical cycles, $\gamma=0.4$). In Fig.~\ref{fig:Xe1p}a the resulting electron momentum distribution corresponding to Xe$^+$ is presented in cylindrical coordinates after integration over the azimuthal angle (see Fig.~\ref{fig:3D_high_energy} for further details). While the symmetric distribution shows similar characteristics to studies performed on other targets such as Ar~\cite{Dura2013, Pullen2014, Wolter2014}, O$_\text{2}$~\cite{Dura2013} and N$_\text{2}$~\cite{Pullen2014}, more low energy features seem to be visible. We find a broad distribution of direct electrons with pronounced low energy features, with the most distinct being the merged VLES and ZES near $p_{||} \approx 0$~a.u. and $p_\perp<0.03$~a.u.. Note that compared to Ar, O$_\text{2}$ and N$_\text{2}$, the ZES in Xe does not appear as clearly (see Fig.~2 of Ref.~\cite{Pullen2014} for a direct comparison). The experimental conditions for Xe, compared to the other species we investigated, included a lower peak intensity of the laser ($I=4\times10^{13}$~W\,cm$^{-2}$ measuring Xe compared to $I=9\times10^{13}$~W\,cm$^{-2}$ measuring Ar~\cite{Wolter2014}) despite a higher extraction field of the ReMi (2.0~V/cm in the case of Xe compared to 1.3~V/cm in the case of Ar~\cite{Wolter2014}). To highlight the individual features three normalized horizontal line-outs are presented in Fig.~\ref{fig:Xe1p}b at values of $p_\perp=0.02$~a.u. (red curve), $p_\perp=0.03$~a.u. (blue curve) and $p_\perp=0.07$~a.u. (orange curve). The $p_\perp=0.03$~a.u. and $p_\perp=0.07$~a.u. curves have been offset by $+0.2$ and $+0.4$, respectively, to enhance visibility. The $p_\perp=0.03$~a.u. curve shows the richest structure with five different structures being resolvable: LES$_\text{1}$ at $p_\parallel\approx0.60-0.70$~a.u., LES$_\text{2}$ at $p_\parallel\approx0.40$~a.u., LES$_\text{3}$ at $p_\parallel\approx0.25$~a.u., LES$_\text{4}$ at $p_\parallel\approx0.15$~a.u. The characteristic V-shape associated with the VLES~\cite{Wu2012} is visible between $\left|p_\parallel\right|\leq0.06$~a.u. The positions of the LES orders (dashed black circles) agree well with predictions~\cite{Kastner2}. For the $p_\perp=0.02$~a.u. and $p_\perp=0.07$~a.u. line-outs the visibility of the various LES orders vary. The other most striking feature is the centre peak ($p_\parallel=0$) for $p_\perp=0.02$~a.u., which corresponds to the ZES. Note that the white dashed line in Fig.~\ref{fig:Xe1p}a represents the acceptance cone for a TOF spectrometer with $\theta_\text{A}=5^\circ$ and with its axis along the polarization direction. Higher transverse momentum parts of the LES orders, the VLES and the ZES, such as those within the grey shaded area in Fig.~\ref{fig:Xe1p}b, are not detected when using such an apparatus. This elucidates the need for 3D imaging of the momentum space in order to observe all low energy recollision features.

\subsubsection{Doubly ionized Xe}
\label{subsec:low_energy_recollisions_double}
\begin{figure}[htb!]
	\includegraphics[width=0.49\textwidth]{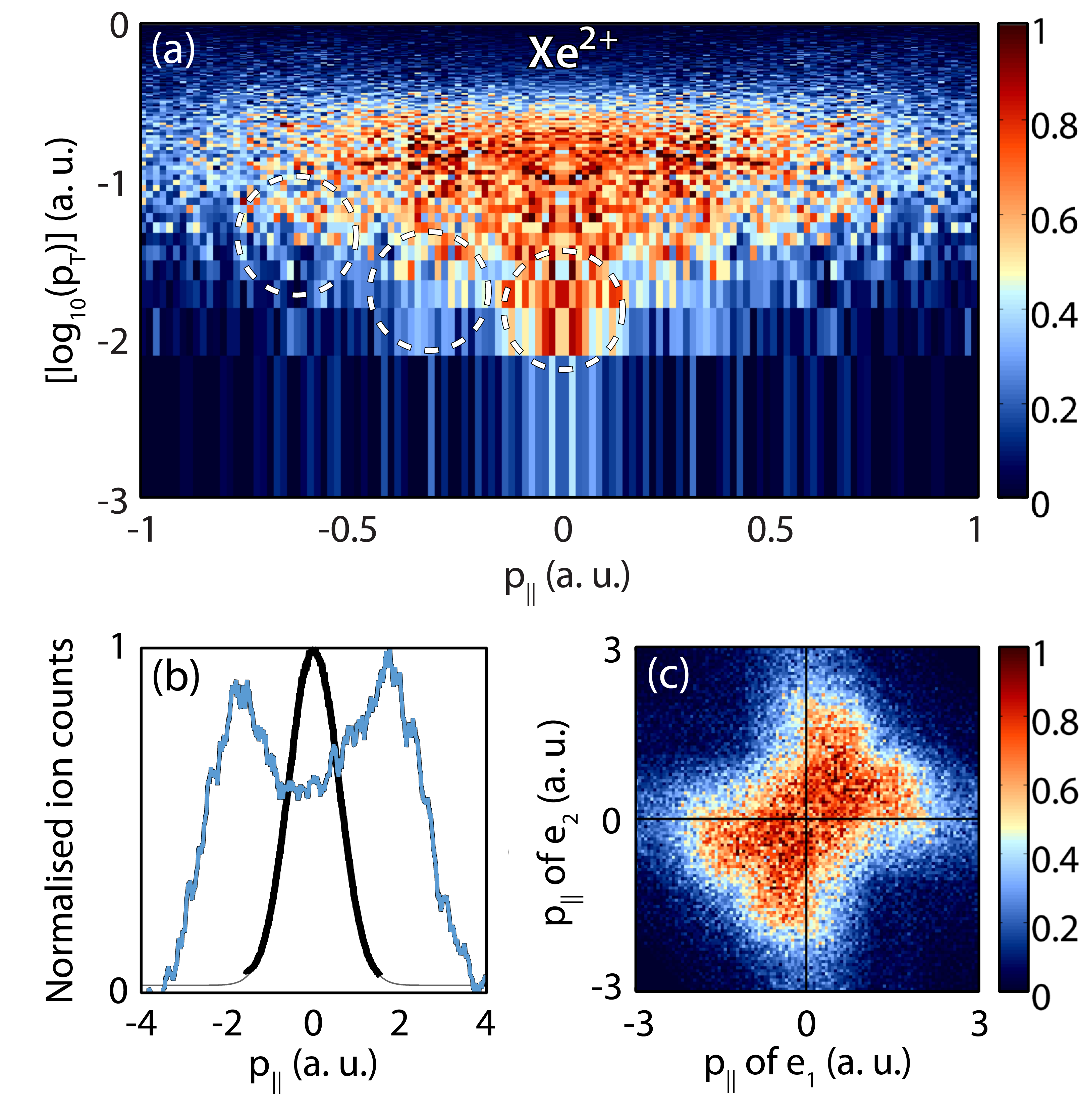}
	\caption{a) The electron momentum distribution corresponding to Xe$^{2+}$ ions for the same experimental conditions as in Fig.~\ref{fig:Xe1p}. b) The longitudinal momentum distribution of the Xe$^{2+}$ ions (blue curve) where a dip at $p_{\parallel}=0$ indicates the involvement of electron recollision. For reference the single peak distribution of the Xe$^+$ ions is also presented (black curve) along with a Gaussian fit (grey curve). c) The correlated two-electron momentum distribution, which features a cross-like shape with a preference to the first and third quadrants.}
	\label{fig:Xe2p}
\end{figure}

Next, we use the coincidence detection capabilities to extract electrons corresponding to doubly ionized xenon atoms Xe$^{2+}$ only. Figure~\ref{fig:Xe2p} shows the first coincidence measurements of doubly ionized Xe with mid-IR fields, and we obtain high measurement statistics due to the stability and repetition rate of our OPCPA system.  The data exhibits a distribution similar to the Xe$^+$ data with three possible structures seemingly emerging above the noise floor: 1) a feature at $p_{\parallel}=\pm\,0.6$~a.u. and $p_{\perp}\approx0.07$~a.u., which is close to LES$_\text{1}$ for Xe$^+$; 2) a feature near $p_{\parallel}=\pm\,0.3$~a.u. and $p_{\perp}\approx0.04$~a.u., which is close to LES$_\text{2}$ or LES$_\text{3}$ for Xe$^+$; and 3) a VLES like distribution at $p_{\parallel}=0$~a.u. and $p_{\perp}\approx0.02$~a.u.. These results indicate, for the first time, that electrons also could undergo multiple recollisions with the ionic Coulomb field during double ionization.

To determine the mechanism behind mid-IR double ionization~\cite{Kopold2000,Feuerstein2001} we plot the parallel momentum distribution of the double ions (see Fig.~\ref{fig:Xe2p}b) and the correlation between the two electrons ((e$_1$,e$_2$), Fig.~\ref{fig:Xe2p}c). The parallel momentum distribution of the double ion, in Fig.~\ref{fig:Xe2p}b, clearly shows the double hump structure which is characteristic of NSDI~\cite{Moshammer2000} and is not explained by sequential double ionization (SDI), where two electrons are emitted independently at subsequent laser field maxima. NSDI exhibits two mechanisms: 1) the recolliding electron can cause the direct emission of a second electron via electron-impact ionization, otherwise known as (e,2e) ionization~\cite{Corkum1993}; or 2) recollision excitation with subsequent ionization (RESI)~\cite{Shaaran2010} where the first electron excites the parent ion which leads to tunnel ionization of a second electron. The form of the double hump in Fig.~\ref{fig:Xe2p}b, i.e. the central dip not being deeply pronounced, is already an indication that both mechanisms of NSDI are present. Further information can be extracted from the electron correlation plot in Fig.~\ref{fig:Xe2p}c. The spectrum is symmetric about $\text{p}_{||,e1}=\text{p}_{||,e2}$ due to the indistinguishability of the two electrons in the experiment. To interpret the measured correlation pattern, we recall that a correlation (first and third quadrants) means that both electrons leave in the same direction (and in the direction opposite to the double ion)~\cite{WeberNature2000}, while an anti-correlation (second and fourth quadrants) means that the electrons escape in opposite directions to each other (the double ion exhibits no momentum)~\cite{Feuerstein2001}. The RESI mechanism was described previously~\cite{Shaaran2010} and due to energy sharing exhibits 'cross-shaped' distributions parallel to the horizontal and vertical axes of the (e$_1$ ,e$_2$) correlation plot. Figure~\ref{fig:Xe2p}c shows features reminiscent of both (e,2e) and RESI. Interestingly, upon first inspection, the shape of the measured correlation shows a striking similarity to Fig.~7c of Ref.~\cite{Bergues2012} where RESI was found to be the dominant mechanism for very different experimental parameters. In that paper, Bergues et al.~\cite{Bergues2012} observed similar features in Ar for near-single-cycle pulses at $\lambda$\,=\,750~nm and at $I=3\times10^{14}$~W\,cm$^{-2}$. Our conditions for Xe are markedly different with a 6.5 cycle pulse (71 fs) at $I=4\times10^{13}$~W\,cm$^{-2}$ in the mid-IR. Without accompanying theoretical calculations it is difficult to determine the exact contribution of either mechanism, however, theoretical investigations into mid-IR double ionization are only now starting to become available~\cite{Tang2012, Zhang2013, Tang2013}. When combined with these first ever full coincidence mid-IR double ionisation results theoretical models will be able to unravel the mechanisms of double ionisation within the mid-IR regime.


\subsection{High energy recollisions}
\label{subsec:high_energy_recollisions}
\begin{figure*}[htb!]
	\includegraphics[width=0.98\textwidth]{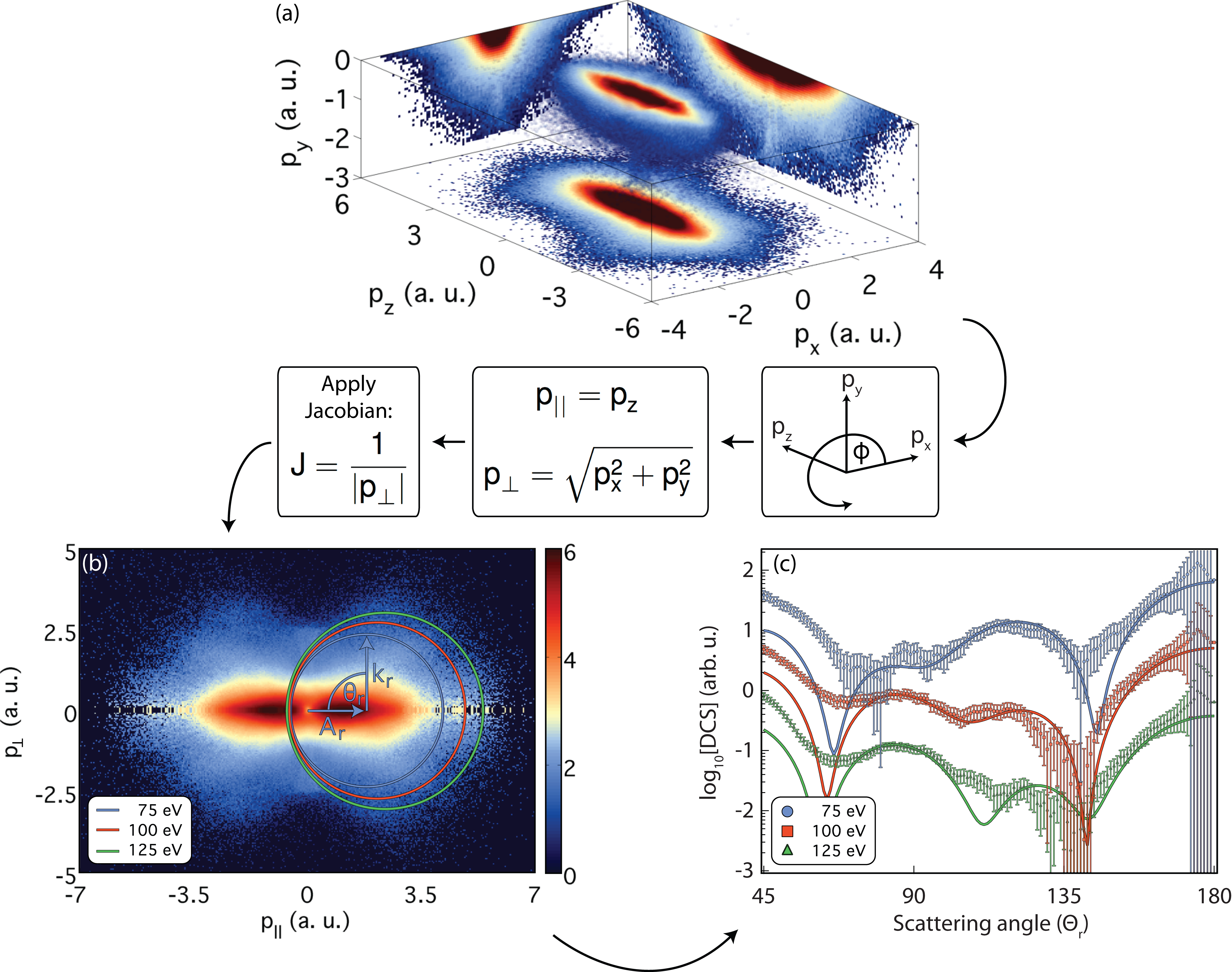}
	\caption{The extraction of elastic DCSs of electrons scattering off Xe atoms using the LIED technique. a) Xe is ionized by the 3.1~$\mu$m radiation with an intensity of $\approx\,6\times10^{13}$~W\,cm$^{-2}$ and the full 3D momentum distribution is detected by the ReMi. b) After integration around the azimuthal angle ($\phi$) and the application of the appropriate co-ordinate transformation Jacobian a 2D representation of the momentum distribution is achieved. c) The DCSs are found by tracing the radius (corresponding to the return electron momentum $k_\text{r}$) of a circle that has been shifted by the vector potential $A_\text{r}$ at the time of electron rescattering. Accurate DCSs are presented for return energies $E_\text{r}$ of 75~eV (blue circles), 100~eV (red squares) and 125~eV (green triangles) over a wide scattering angle range. The solid lines are taken from the NIST database~\cite{NIST}.}
	\label{fig:3D_high_energy}
\end{figure*}

The capability of our apparatus towards investigating SFI at high electron energies is demonstrated next. We measure doubly differential electron scattering cross-sections (DCSs) using LIED~\cite{Zuo1996,Spanner2004,Yurchenko2004}. The realization of LIED is dependent on two key points: 1) the generation of high impact energies for the recolliding electrons; and 2) the ability to extract field free DCSs despite the presence of the strong laser field. Both conditions are fulfilled with our apparatus. Previously, such measurements of the DCS were either limited due to the shorter wavelength of the Ti:Sapphire laser, or the inability of collecting the full 3D momentum distribution~\cite{Ray2008, Okunishi2008, Okunishi2011, Blaga2012, Xu2012, Okunishi2014}. Here, we show the first LIED experiment based on 3D momentum detection in full particle coincidence.

Mid-IR LIED is achieved by operating the ReMi at high static electric and magnetic fields to detect the entire 3D momentum distribution of the emitted electrons. Figure~\ref{fig:3D_high_energy}a displays the momentum distribution created during the interaction of our intense 3.1~$\mu$m radiation with Xe atoms. After an integration over the azimuthal angle $\phi$ and the application of the appropriate co-ordinate transformation (Jacobian $J=1/|p_\perp|$), a 2D momentum map representation of the detected momentum distribution is extracted (Fig.~\ref{fig:3D_high_energy}b). The momentum map covers a six order of magnitude range in counts and features an unprecedentedly rich structure. DCSs are extracted from the 2D momentum distribution (or diffraction pattern) in a manner similar to conventional electron diffraction (CED) provided that three critical criteria are satisfied: 1) QS conditions must be achieved so that the final momentum distribution of electrons can be expressed as the product of the field-free DCS multiplied by the returning electron wave packet~\cite{Chen2009}. In this case, the vector potential of the driving field can easily be calculated and taken into account and field-free DCSs can be extracted; 2) the returning electron wavepacket must be energetic enough to probe the core of the target and not the valence electronic structure (i.e. $E_\text{r} \geq 50$~eV); and 3) the recolliding electron must propagate sufficiently far away from the interaction region such that it approximates a plane wave upon return. These three requirements can only be simultaneously satisfied by intense mid-IR sources ($I\geq5\times10^{13}$~W~cm$^{-2}$ and $\lambda \geq 2~\mu$m). Once these stringent conditions are satisfied, the achieved momentum transfers are similar between LIED and CED despite the difference in incident electron energies and  scattering angles. That means in mid-IR LIED, electron energies $E \geq 50$~eV that scatter at angles between $30^\circ<\theta_\text{r}<180^\circ$ are typically utilized while in CED the electron energies are on the order of $10^4-10^5$~eV and scatter at forward angles ($0^\circ<\theta_\text{r}<10^\circ$). This means that the classical momentum transfer ($q=2 k_\text{r} \sin{\theta_\text{r}/2}$, where $k_\text{r}$ is the electron incident momentum) is comparable for both methods and, as such, mid-IR strong-field induced re-scattered electrons contain structural information of the target gas.

The theoretical framework behind the analysis of the mid-IR LIED pattern is provided by the quantitative re-scattering (QRS) theory~\cite{Chen2009,Lin2010,Xu2010}. This model states that angularly resolved DCSs (up to an overall scaling factor) can be extracted from the measured electron momentum distribution by tracing the measured counts along the circumference of a circle of radius $k_\text{r}$ that has been shifted from the origin by the vector potential of the laser at the time of re-scattering ($A_\text{r}$). Here $k_\text{r} = \sqrt{2E_\text{r}}$ is the incident momentum of the re-scattering electron with energy $E_\text{r}$. The relation between the incident energy of the returning electron $E_\text{r}$ and the final electron energy of the rescattered electron $E_{\text{kin}}$ is plotted in the inset of Fig.~\ref{fig:keldysh} for both long and short trajectories (derivation from e.g. in section II.A. of \cite{Chen2009}). An example of this extraction procedure for 75 eV (blue), 100 eV (red) and 125 eV (green) energy electrons is presented in Fig.~\ref{fig:3D_high_energy}b. The extracted DCSs are presented in Fig.~\ref{fig:3D_high_energy}c as a function of the scattering angle. As each returning electron encounters a slightly different vector potential, the rescattering circles are shifted by differing amounts. The 75 eV (blue circles) and 125 eV (green triangles) data have been scaled by factors of $10^{1}$ and $10^{-1}$ compared to the 100 eV (red circles) data, respectively, in order to enhance visualization. The error bars are derived from Poissonian statistics and are propagated through the integration and co-ordinate transformation procedures. The large error bars at $\theta_\text{r}\approx\,180^\circ$ are due to the application of the Jacobian mentioned above. Also presented in Fig.~\ref{fig:3D_high_energy}c are the Xe DCSs provided by NIST~\cite{NIST}, which have been scaled to overlap with the experimental data. Our data are able to accurately resolve most of the predicted rich structure for $\theta_\text{r}\geq60^\circ$. For these angles the maxima and minima as well as slope changes are generally well reproduced while for smaller scattering angles ($\theta_\text{r}<60^\circ$) the data is 'contaminated' through background arising from direct electrons. While these direct electrons cause a slight mismatch between the experimental and theoretical data, the minima between $\theta_\text{r}=50-60^\circ$ are still resolvable. Overall we observe an excellent match between our measured data and the theoretical expectations. We note that the presented NIST data is calculated for neutral Xe. It has been shown that the electron energies utilized here ($E_{\text{kin}}$ $>$ 50\,eV) are core penetrating and can be used to extract structural information~\cite{Pullen2014a}. Therefore the electronic structure of the target has a negligible effect on the extracted DCSs for these experimental conditions. These results show that field free DCSs can indeed be extracted over a wide range of energies and scattering angles at unprecedented accuracies.

\section{Summary and outlook}
\label{sec:summary}

We have presented a unique methodology to investigate mid-IR driven SFP with a high repetition rate mid-IR OPCPA system in combination with a ReMi that is able to detect ionization fragments over a six order of magnitude energy range. This apparatus now permits unambiguous investigations of SFI and related effects in the deep QS regime, thereby allowing experimental observations to be scrutinized against theoretical descriptions. We demonstrate our concept across the entire accessible kinetic energy scale by investigating SFI and electron recollision with Xe atoms. We observe low energy structures up to the fourth order, the VLES and the ZES in Xe but find a much reduced ZES that would arise from capture into Rydberg states and subsequent ReMi field ionization. This surprising absence of a strong ZES peak in Xe already warrants further investigation as it will prompt information about the FTI process. Next, first coincidence measurements of mid-IR driven NSDI are shown. The analysis of the data hints at the combined action of (e,2e) and RESI mechanisms. Interestingly, this very first (e1,e2) correlation map in the deep QS regime ($\gamma = 0.4$) for a 6.5 cycle pulse exhibits features that were observed for entirely different conditions with a near-single-cycle CEP-stable pulses at $\lambda$\,=\,750\,nm in Ar and for $\gamma = 0.7$~\cite{Bergues2012}. Lastly, we exploit the high energy detection capabilities of our apparatus with mid-IR LIED. We achieve core penetrating conditions that permit extraction of the doubly differential elastic electron scattering cross section in Xe and the retrieved DCSs match published NIST data very well. Probing atomic structure with mid-IR LIED carries clear potential for the realization of femtosecond time resolved imaging of molecular structure and for tracking atomic constituents during structural changes such as isomerization or dissociation. In fact, recent results from our group show that the method can indeed be used to determine the entire conformation of polyatomic acetylene (C$_\text{2}$H$_\text{2}$)~\cite{Pullen2014a}. 

\section{Acknowledgements}
We acknowledge support from the Spanish Ministerio De Economia Y Competitividad (MINECO) through ``Plan Nacional'' (FIS2011-30465-C02-01), the Catalan Agencia de Gesti\'{o} d'Ajuts Universitaris i de Recerca (AGAUR) with SGR 2014-2016. This research has been supported by Fundaci\'{o} Cellex Barcelona, LASERLAB-EUROPE grant agreement 228334 and COST Action MP1203. B. W. was supported by an AGAUR fellowship (FI-DGR 2013-2015). M.G.P is supported by the ICFONEST+ programme, partially funded by the Marie Curie Co-funding of Regional, National and International Programmes -- COFUND (FP7-PEOPLE-2013-COFUND) action of the European Commission, the ``Severo Ochoa'' Program of the Spanish Ministry of Economy and Competitiveness, and ICFO.\\ \\B.W. and M.G.P. contributed equally to this work.


\begin{thebibliography}{30}
\bibitem{footnote1} The $^{129}$Xe$^{2+}$ peak is composed of two peaks due to the NSDI mechanism (polarization along the static electric field of the ReMi) causing one recoil-ion to go straight towards the detector and the other to initially go away from the detector.
\end{thebibliography}
\end{document}